\documentclass[11pt]{article}

\begin{document}

\title{Quantum cryptography: a practical information security
  perspective%
\thanks{This paper was originally posted with the title
  {\it Why quantum cryptography?} It was published in
Quantum Communication and Security, Proceedings, NATO Advanced Research
Workshop, edited by M. \.Zukowski, S. Kilin and J. Kowalik, p.~175--180
(IOS Press, Amsterdam, 2007).}}

\author{Kenneth G. Paterson, Fred
  Piper and R\"udiger Schack \\ {\it Department of Mathematics, Royal
    Holloway, University of London} \\ {\it Egham, Surrey TW20 0EX,
    UK}}

\date{16 December 2004}
\maketitle

\abstract{
  Quantum Key Exchange (QKE, also known as Quantum Key Distribution or QKD)
  allows communicating parties to securely establish cryptographic keys. It is
  a well-established fact that all QKE protocols require that the parties have
  access to an authentic channel. Without this authenticated link, QKE is
  vulnerable to man-in-the-middle attacks. Overlooking this fact results 
  in exaggerated claims and/or false
  expectations about the potential impact of QKE. In this paper we present a
  systematic comparison of QKE with traditional key establishment protocols in
  realistic secure communication systems.}

\section{Introduction}

It is impossible to obtain information about a physical system without
disturbing it in a random, uncontrollable way. This fundamental
quantum-mechanical law guarantees the security of QKE protocols by enabling
the communicating parties to put an upper bound on how much an eavesdropper
can know about the key. QKE protocols such as BB84 \cite{Bennett1984} have
been proved to be secure under the assumption that the known laws of quantum
physics hold \cite{Mayers2001}. Given this assumption, QKE is secure even in
the presence of an adversary with unlimited computational power.
See~\cite{Gottesman-0111} for an overview of QKE and other aspects of quantum
cryptology. Following common usage, we will
call {\em unconditionally secure\/} any protocol whose security does not
depend on assumptions about the computational power of a potential adversary.

Although QKE requires the use of (currently expensive)
special purpose hardware and/or networks, secure communications
systems based on QKE appear to enjoy an advantage over most
systems based on public key cryptography. For the latter would
become insecure if progress in algorithms for integer
factorization or discrete logarithms were made, and in particular
if a quantum computer were built \cite{Gisin2002A}. Because of
this, unconditionally secure QKE is often portrayed as being the
ideal solution to the problem of distributing cryptographic keys.
We will now show that this view only tells part of the story and
has led to exaggerated claims and/or false
expectations about the advantages of systems using QKE (e.g., see 
\cite{Klarreich2002}).

At least two other components are required in addition to the
basic QKE protocol in order to make a secure communications
system.

Firstly, QKE requires that the parties have access to an {\it
authentic\/} channel (which need neither be quantum nor secret).
Any QKE protocol that does not fulfill this requirement is
vulnerable to a man-in-the-middle attack. For previous discussions
of authentication in QKE, see~\cite{Bennett1992,Dusek1999a}. 
The authentication mechanism
used to provide the authentic channel may or may not be
unconditionally secure. We refer to the combination of
authentication mechanism and QKE protocol as the {\em key exchange
sub-system}. The key exchange sub-system will only be
unconditionally secure if the authentication mechanism and QKE
protocol are.

Secondly, the keys that are exchanged in the QKE protocol are in turn used to
protect communication of data using an encryption algorithm. The encryption
algorithm may or may not be unconditionally secure. The {\em overall\/}
QKE-based communications system will only be unconditionally secure if the
authentication mechanism, QKE protocol and encryption algorithm all are.

It is common for vendors of QKE-based systems to offer the option of
combining an
unconditionally secure key exchange sub-system with a conventional
encryption algorithm such as 3DES or AES \cite{idquantique,magiq}.
Of course, an overall communications system constructed in this
way cannot be unconditionally secure. We call such systems {\em
hybrid} systems.

Typically, a QKE protocol forms only one component of a
complete communications system. Such a system can (in general)
only be as secure as its weakest component. Thus, in assessing the
security offered by a system using QKE, one must examine the
entire system and not rely just on a claim of unconditional
security for the QKE protocol component.

These issues appear to be well-known in the quantum cryptography community. 
Yet to our knowledge there has been no systematic analysis, from the point of
view of practical information security, of how these issues impact on the
applicability of QKE. Similarly, little has been done to examine how QKE 
compares to more traditional approaches to establishing secure communications
in terms of practicality, cost, and security levels (both those offered by the
different approaches and those actually needed in applications). 

The present paper intends to provide such an analysis. Our analysis is driven
by an examination of the need to provide an authentication channel in QKE
systems. In Sections 2 and 3, we show that an
unconditionally secure key exchange sub-system making use of QKE
requires the pre-establishment of a symmetric key between the
communicating parties. We then examine the practical consequences of this in
Section 4. 

In what follows, we make a division
between systems using {\em public-key authentication\/} and
systems using pre-established {\em symmetric keys\/} for
authentication. Furthermore, among the latter systems, we will
distinguish between hybrid systems and unconditionally secure
systems.

\section{Systems using public key authentication}

In such
systems, public key cryptographic mechanisms, e.g., digital
signatures, are used to provide the authentic channel needed for
QKE. The key exchange sub-system, and hence the overall
communications system, will be no more secure than the public key
authentication mechanism on which it is based. For example, if RSA
digital signatures are used for authentication, a system of this
type would become insecure if quantum computers became available.
Hence such a system does not offer unconditional security.
Moreover, any system using QKE requires a
quantum channel (e.g., an optical fiber) between the communicating
parties. Commercial QKE products can use existing telecom fiber
optics networks to provide the quantum channel
\cite{idquantique,magiq}.

Nevertheless, a system of this type may still offer some security
advantages over traditional (i.e.\ non QKE-based) approaches. In
particular, in any successful attack on such a system, the public
key authentication mechanism would have to be broken before or
during the execution of the QKE protocol. This is in contrast to a
system using only classical information and traditional
key-establishment techniques, where the messages exchanged in
order to establish a key can be stored by the adversary and
analyzed at some point in the future, possibly using more advanced
cryptanalytic techniques than are available at the time of key
establishment.

It follows from the above that, if the authentication mechanism is
unbroken at the time of key establishment, and if the one-time pad
is used as the encryption algorithm, then transmitted data remain
secure indefinitely.  Thus, in order to guarantee the long-term
security of communications, one would only need to be concerned
about the capabilities of attackers today rather than in the
future. This could be an attractive solution for protecting
government secrets, for example. Similarly, if the authentication
mechanism is unbroken at the time of key establishment, and if an
encryption algorithm such as AES is used, then the data remain
secure as long as that encryption algorithm remains secure. Such a
system would also be resilient to attacks in which an adversary
was able to learn the private keys of the communicating parties
and then mounted a passive eavesdropping attack on subsequent
exchanges. Naturally, such a system would not resist active
attacks subsequent to private key compromise.

It should be mentioned that there exist proposals for quantum public
key protocols, where the quantum state of a string of qubits (quantum
bits) is used as a key \cite{Gottesman-0105a}.  Storage, distribution
and manipulation of these quantum keys, however, require quantum
information processing capabilities beyond the reach of current
technology. Using public quantum keys for authentication is thus not
an option now or in the foreseeable future.

\section{Systems using symmetric key authentication}

If
the communicating parties already share a secret, symmetric key,
then they can use that key to establish the authentic channel
needed to support QKE. In essence, both parties attach
cryptographic tags to their messages on that channel, the tags
depending both on the message transmitted and on the shared key.
Such an authentication mechanism can offer either conditional or
unconditional security.

The classic approach to providing an unconditionally secure
authentic channel is to make use of a message authentication code
(MAC) due to Wegman and Carter \cite{Wegman1981}. In this
approach, the parties use the Wegman-Carter MAC together with the
pre-established key to authenticate all their messages. The key
can be much shorter than the messages being authenticated. All
currently existing authentication schemes which offer
unconditional security are similar to the Wegman-Carter approach
in that they depend on a pre-established symmetric key.

A key exchange system using QKE and symmetric key authentication
differs from a traditional key exchange system using public key cryptography
in two main respects. Firstly, as we have already mentioned, it
requires a quantum channel between the communicating parties.
Secondly, it requires the initial establishment and management of
secret keys between the communicating parties. This is certainly
feasible, even on a large scale; a good example is provided by GSM
mobile communications systems. In a GSM system, a user's
Subscriber Identity Module (SIM) contains a 128 bit symmetric key
which is shared with the subscriber's network service provider.
This key is used in an authentication protocol, one product of
which is a fresh, symmetric data encryption key. In mid-2003, GSM
systems were in operation in 205 countries, with more than 1
billion subscribers \cite{GSMworld}. Such symmetric hierarchical
systems pre-date the advent of public key cryptography and have a
long and successful history of use in telecommunications and
finance.

We now further subdivide our study of systems using pre-established
symmetric keys for authentication.

\subsection{Hybrid systems}

QKE can be used as a component
in a hybrid system, where the secret bits resulting from the QKE
protocol are used as keying material in a symmetric
encryption algorithm such as 3DES or AES.

Such a hybrid communications system using QKE can offer security
advantages over conventional alternatives. For example, it may
provide unconditionally secure refresh of cryptographic keys if an
unconditionally secure authentication mechanism is used. However,
the security of the overall communications system will be limited
by the security of the symmetric encryption algorithm used. The
overall security offered by this approach is therefore only
conditional.

\subsection{Unconditionally secure systems}

Here the
communicating parties must establish an authentic channel with
unconditional security and use an unconditionally secure
encryption algorithm. An unconditionally secure encryption
algorithm is provided only by the one-time pad. In order to achieve
this level of security for encryption, as many key bits as there
are message bits must be established by the QKE protocol. This may
be a problem in some practical applications, as the key bit rates
of current QKE systems are relatively small. In a traditional
one-time pad system (not making use of QKE), the pre-established
key must be at least as long as the data to be communicated. A QKE
system has an advantage here in that the pre-established key can
be relatively short, as it is used only to authenticate an initial
run of the QKE protocol, with part of the keying material
exchanged in that run being used to authenticate subsequent runs.

To summarize, by combining an unconditionally secure
authentication scheme with a QKE protocol, one can produce a key
exchange sub-system which enjoys a level of security that can be
established unconditionally, assuming only the validity of the
laws of quantum physics. If the one-time pad is used as the
encryption algorithm, then the overall communications system can
also be made unconditionally secure.

\section{Discussion}

It is likely that using QKE with public key authentication (and
therefore not requiring pre-establishment of a symmetric key) has
security benefits when the long-term security of data is of
importance. There may also be some security advantages in using QKE
in hybrid systems as described above.

However, QKE loses much of its appeal in these settings, as the overall system
security is no longer guaranteed by the laws of quantum physics alone. To
obtain an overall communication system with unconditional security, an
unconditionally secure key exchange sub-system is required. From our analysis,
it is evident that to obtain such a sub-system, a pre-established
secret key is required. We note that this requirement is seldom emphasized by
proponents of QKE. It is now also clear that QKE, when unconditionally secure,
does not solve the problem of key distribution.  Rather, it exacerbates it, by
making the pre-establishment of symmetric keys a requirement. The often-made
comparison between the unconditional security of QKE and the conditional
security offered by public key cryptography overlooks this requirement of QKE.
The establishment and subsequent management of symmetric keys is a significant
undertaking, and any comparison of QKE and public key cryptography should take
this fact into account.

The pre-established symmetric keys needed to provide authentication in
an unconditionally secure QKE protocol could instead be used directly
in a symmetric encryption algorithm, or as the basis for a symmetric
hierarchical system like that employed in GSM and many other systems.  
Thus a complete evaluation of the purported
benefits of QKE should also compare the level of security offered by
QKE to the level that can be achieved using conventional symmetric
techniques alone.

For a well-designed symmetric encryption algorithm, the best
attack should require the attacker to expend an amount of effort
equivalent to that of an exhaustive key search in order to break
the algorithm, even if large amounts of plaintext and ciphertext are
available to the attacker. With the key lengths available today in algorithms
like AES, an exhaustive key search is simply not a realistic attack.
Furthermore, all known attacks against such algorithms using
quantum computers would be easily countered simply by doubling the
key length. Thus the only applications where using an
unconditionally secure QKE protocol appears justified are those
for which the level of security offered by the best available
symmetric encryption algorithm is judged insufficient because of
the risk that the algorithm turns out {\em not\/} to be
well-designed and there are advances made in the cryptanalysis of
that algorithm. In such applications, the QKE protocol should only
be used with the one-time pad for encryption, since any advance in
cryptanalysis of symmetric algorithms may also compromise the
encryption algorithm used in a hybrid QKE system. We suggest that
this set of applications is in fact rather limited: we do not
foresee many commercial uses where the expense associated with
such a degree of security would be warranted. Adding to this the
fact that conventional techniques have no requirements for
special-purpose hardware or dedicated networks, we believe that
the traditional symmetric approach has much to offer in comparison
with unconditionally secure QKE. Whilst it is certainly worthwhile
to study the impact that the advent of quantum computing might
have on conventional cryptography, it is not true that large-scale
quantum computing would bring about the death of all conventional
cryptographic approaches. Rather, it would serve to enhance the
value of long-established symmetric key management techniques.

\section*{Acknowledgment}

We thank Burt Kaliski for useful discussions on the content of this paper.


\end{document}